\documentclass[USenglish, reprint, superscriptaddress,
 amsmath,amssymb,
 aps,
 longbibliography,
 footinbib,
]{article}	

\usepackage[utf8]{inputenc}				
\usepackage[big,online]{dgruyter}	
\usepackage{lmodern} 
\usepackage{microtype}
\usepackage[numbers,square,sort&compress]{natbib}
\usepackage{graphicx}
\usepackage{dcolumn}
\usepackage{bm}
\usepackage[dvipsnames]{xcolor}
\usepackage{comment}
\usepackage{float}
\usepackage[normalem]{ulem}
\usepackage[square,numbers]{natbib}
\theoremstyle{dgthm}

\theoremstyle{dgdef}

\newcommand{\dwt}{\Delta\tilde\omega}
\newcommand{\dw}{\Delta\omega}
\newcommand{\gnr}{\gamma_{\text{nr}}}
\newcommand{\gr}{\gamma_{\text{r}}}
\begin{document}

	\articletype{Research Article}
	\received{Month	DD, YYYY}
	\revised{Month	DD, YYYY}
  \accepted{Month	DD, YYYY}
  \journalname{De~Gruyter~Journal}
  \journalyear{YYYY}
  \journalvolume{XX}
  \journalissue{X}
  \startpage{1}
  \aop
  \DOI{10.1515/sample-YYYY-XXXX}

\title{ Nonlinear optical heating of all-dielectric super-cavity: \\  efficient light-to-heat conversion through giant thermorefractive bistability}
\runningtitle{Nonlinear optical heating}

\author{Daniil Ryabov, Olesiya Pashina, George Zograf, Sergey Makarov, Mihail Petrov*}
\affil{\protect\raggedright 
ITMO University, Department of Physics, Saint Petersburg, Russia}

	
\abstract{Optical heating of resonant nanostructures is one of the key issues in modern nanophotonics, being either harmful or desirable effect depending on the applications. Despite a linear regime of light-to-heat conversion being well-studied both for metal and semiconductor resonant systems {is} generalized as {a} critical coupling condition, the clear strategy to optimize optical heating upon high-intensity light irradiation is still missing. This work proposes a simple analytical model for such a problem, taking into account material properties changes caused by the heating. It allows us to derive a new general critical coupling condition for the nonlinear case, requiring a counterintuitive initial spectral mismatch between the pumping light frequency and the resonant one. Based on the suggested strategy, we develop an optimized design for efficient nonlinear optical heating, which employs a cylindrical nanoparticle supporting the quasi bound state in the continuum mode (quasi-BIC or so-called 'super-cavity mode') excited by the incident azimuthal vector beam. Our approach provides a background for various nonlinear experiments related to optical heating and bistability, where self-action of the intense laser beam can change resonant properties of the irradiated nanostructure.}


\maketitle

\section{Introduction} 

The strong resonant response of all-dielectric resonant nanosystems in the visible and infrared region along with the diversity of their optical properties opens the way for various applications in nonlinear and laser optics~\cite{kuznetsov2016optically}. Remarkably, such low-loss nanostructures can support high-Q modes under proper management of radiative losses~\cite{Koshelev2019} even in single nanocavities on dielectric substrates owing to excitation of super-cavity modes~\cite{koshelev2020subwavelength}, or in the one- and two-dimensional arrays of resonant nanostructures~\cite{Bulgakov2019, Bulgakov2017a, Kornovan2021, Azzam2021}, which are often related to the bound state in the continuum~\cite{hsu2016bound} (BIC) or quasi-BIC in the systems of finite size. Even though in many applications, the inherent optical heating of resonators {is} considered {as} a parasitic effect, strong thermooptical coefficients of all-dielectric materials can be utilized for developing thermally tunable and reconfigurable nanophotonic devices~\cite{makarov2017light}. Indeed, the achievements of thermo-nanophotonics based on all-dielectric~\cite{zograf2021all} nanostructures (similarly to thermoplasmonics~\cite{govorov2007generating,baffou2013thermo,baffou2020applications}) showed that efficient nano- and microscale sources of heat are in strong demand in various nanoscience applications.

From this point of view, the problem of efficient heating of all-dielectric nanostructures requires special optimization, depending on the final application of the nanophotonic design.  Based on this approach and employing advanced methods of nanothermometry, the case of linear optical heating of all-dielectric Mie-resonant nanoparticles was successfully described theoretically~\cite{Danesi2018,Rocco2021, Pashina2022Feb} and demonstrated experimentally~\cite{zograf2017resonant,Aouassa2017, Celebrano2021}. However, the situation becomes much more complicated once the optical nonlinearity due to thermorefractive effects is taken into account~\cite{Jiang2020}. Recently, there has been a significant progress in the field of nonlinear thermal nanophotonics with resonant all-dielectric  \cite{Tang2021} and plasmonic \cite{Sivan2017} systems, showing large values of thermal nonlinearity in single resonant nanostructures \cite{Li2021,Huang2022,Duh2020,Zhang2020a}.    In this prospective, the problem of efficient heating is tightly  connected to maximizing the absorption of resonators~\cite{Grigoriev2015,Miroshnichenko2018}. While {the} critical coupling condition is required for the most optimal linear heating regime~\cite{zograf2021all}, the exact conditions of maximal optical heating in the nonlinear regime, when the elevated temperature drives change of the real and imaginary parts of the refractive index, are yet to be identified.

\begin{figure}[t]
\begin{centering}
\includegraphics[width=0.65\columnwidth]{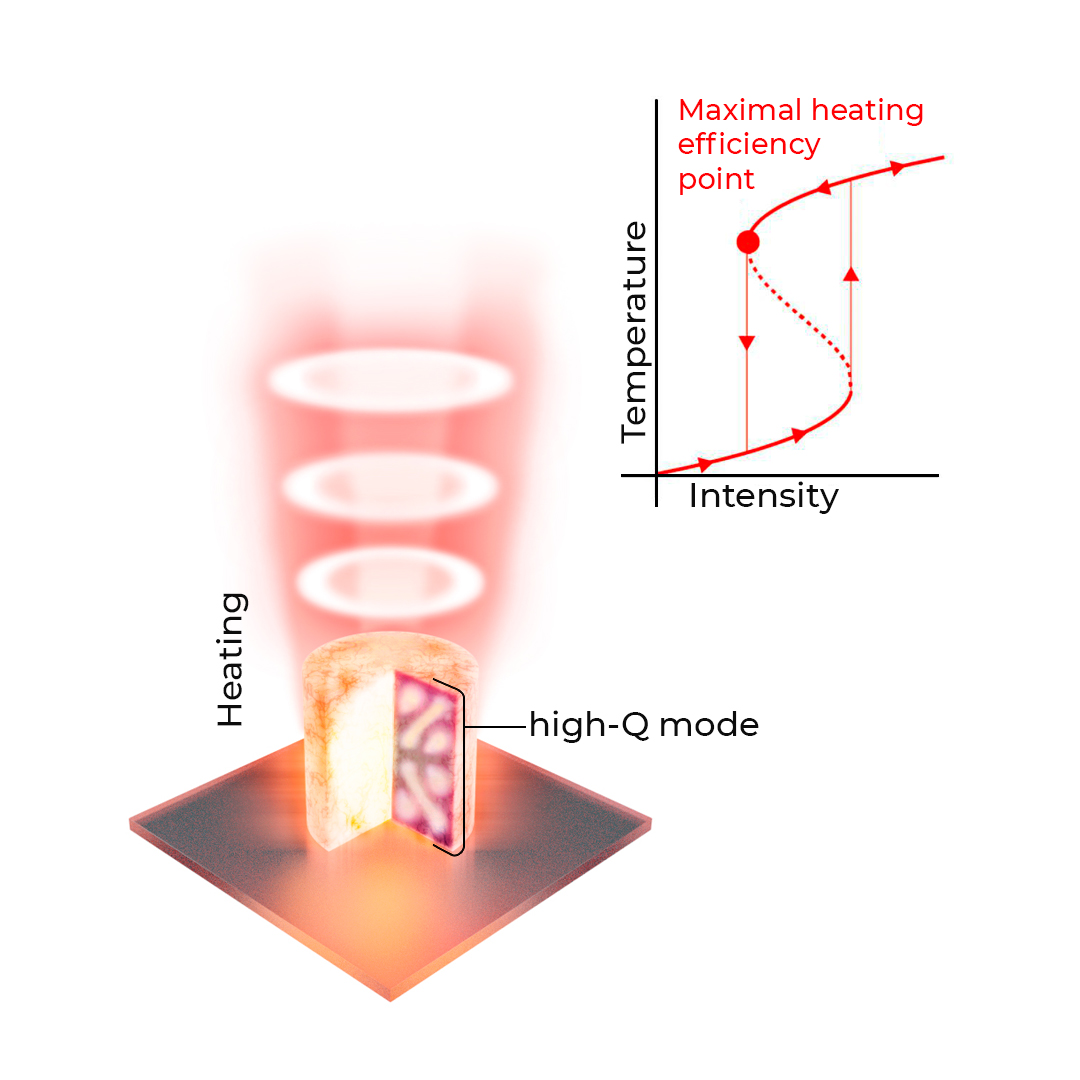}
\caption{\label{fig:general} A schematic illustration of the main idea: optical heating of the nanoresonator leads to a spectral shift of the mode frequency inducing strong thermooptical nonlinearity. As the result, the bistable state can be achieved allowing for further efficient heating in the nonlinear critical coupling regime. }
\end{centering}
\end{figure}


In this work, we firstly develop a simple analytical model for the nonlinear optical heating of a single-mode resonator supporting efficient light-to-heat conversion and bistable regime of operation, and then employ numerical simulations to propose a realistic design based on the super-cavity. The developed formalism and numerical design dealing with doped silicon cylindrical particles (see Fig.~\ref{fig:general}) allow for the creation of novel thermooptical nanophotonic devices for optical switching and signal processing, as well as useful for the experiments where heat generation in resonant nanostructures should be minimized.

\section{Optical Heating of a nanoresonator}
\label{sec:the_model}
Optical heating of matter is a rather complex process, which combines free carrier generation, their interaction with light and phonons, and transport of the phonons across the material~\cite{Cunha2020}. However, in the steady-state continuous wave (CW) excitation of subwavelength resonators considered in this paper the situation becomes much simpler. For a large class of nanophotonic structures such as nanoparticles and nanoantennas made of material with high thermal conductivity (metals or semiconductors) as compared to surrounding material (air, glass, water) the steady-state temperature appears to be homogeneously distributed along the nanostructure~\cite{Baffou2017} and can be described by the expression~\cite{baffou2013thermo} 
\begin{align}
\label{eq:temp}
\Delta T=C\dfrac{P_{\text{abs}}}{\eta_{\text{eff}} R}, 
\end{align}
\noindent
where $P_{\text{abs}}$ is the light power absorbed inside the resonator due non-radiative (ohmic) losses, $R$ is the typical radius of the sphere of the same volume as the nanostructure,  $\eta_{\text{eff}}$ is the effective thermal conductivity of surrounding media, and $C$ is a constant which is defined by the shape of the nanoantenna and geometry of the problem. 

The problem of efficient heating of nanoresonator is, thus, reduced to maximization of the absorbed power, which has its fundamental limitation~\cite{Miroshnichenko2018}. This limit,  however, can be reached in the {\it critical coupling } regime when scattered and absorbed power are balanced, which paves the way for optimizing the absorption losses~\cite{Grigoriev2015}. However, despite the critical coupling condition is usually considered for a linear system, it still holds in strongly nonlinear Kerr systems~\cite{NireekshanReddy2013} or two-level systems with saturated absorption~\cite{Longhi2011a,Shen2010} allowing for coherent perfect absorption. Here, we discuss the nonlinear critical coupling in the context of optical heating of single nanoresonators. We start by considering a single-mode nonlinear resonator within a coupled-mode theory~\cite{Suh2004} as a toy model to identify the main conditions to reach the maximal absorption condition.      

\begin{figure}[t]
\includegraphics[width=\textwidth]{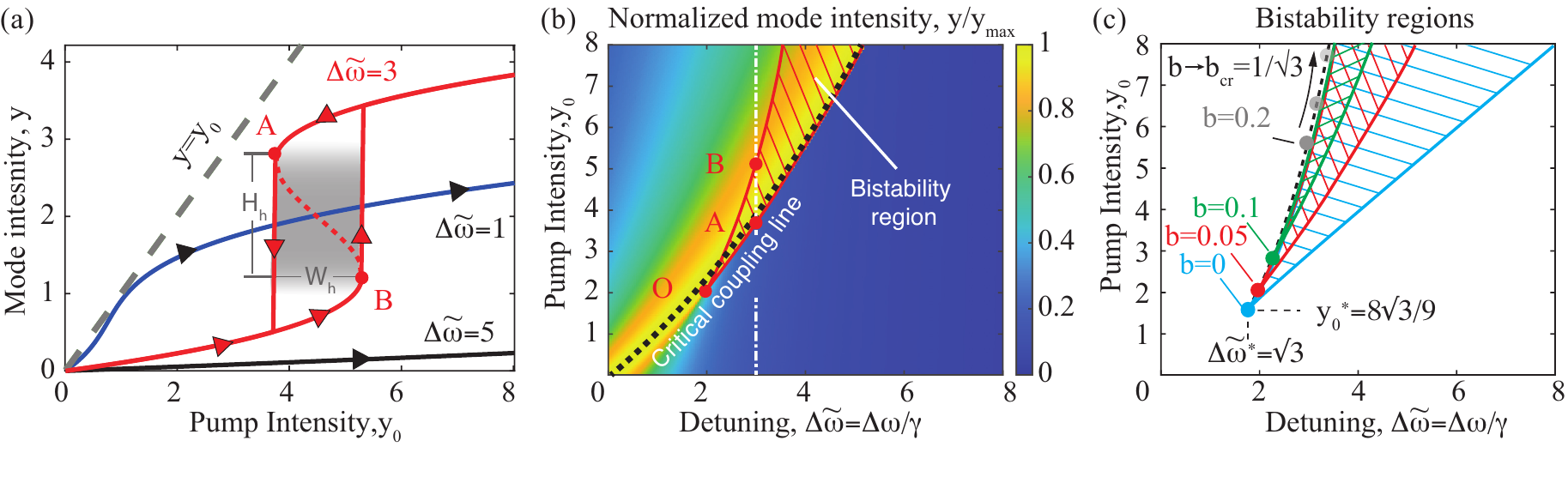}
\caption{(a) {Dimensionless} mode intensity {$y = \alpha|a|^2/\gamma$} as a function of {dimensionless} pumping intensity {$y_0 = \alpha\gamma_r|f|^2/\gamma^3$} for different detuning parameters $\dwt=1,3,5$ and relative nonlinear cefficient $b=0.05$. {Gray region corresponds to hysteresis area with height $H_{\text{h}}$ and width $W_{\text{h}}$}. (b) The map of the heating efficiency as the function of the detuning and the pump intensity for the upper branch of the solution $b=0.05$.  The bistability region is shown with a hatch. The critical coupling line (maximal heating efficiency) is denoted with the dashed line. (c) The evolution of the bistability regions with the variation of relative nonlinear coefficient $b$.   \label{fig:Model_Bistab}}
\end{figure}
\subsection{Critical coupling in linear regime}
We consider a subwavelength  resonator with resonance at the frequency $\omega_0$  having radiative and non-radiative (ohmic) losses rates $\gamma_{\text{r}}$ and $\gamma_{\text{nr}}$ correspondingly.  The latter are responsible for light absorption inside the resonator  and its consequent heating. Indeed, the absorbed power  $P_{\text{abs}}$ is proportional to non-radiative losses  $\gamma_{\text{nr}}$ and the total  electric energy $W$ stored inside the resonator: $P_{\text{abs}}=2\gamma_{\text{nr}}W$.  A  single mode resonator can be well-described by the temporal coupled mode theory~\cite{Suh2004} and the amplitude $\tilde a$ of the excited mode can be given by the equation~\cite{haus1984waves}:
\begin{equation} 
    \frac{d\tilde{a}}{dt} = (-i\omega_0 + \gamma)\tilde{a} + i\sqrt{\gr} \tilde{f}, \label{eq:amp_init}
\end{equation}
 where $\omega_0$ is the eigenfrequency of the resonator, $\gamma=\gr+\gnr$ is the total loss rate, $\tilde{f}$ corresponds the amplitude of the incident wave. In the stationary regime under harmonic excitation  $\tilde{f} = f \exp(-i\omega t)$, the spectral amplitude intensity of the field  $\tilde{a} = a \exp(-i\omega t)$ can be found from Eq.\eqref{eq:amp_init}: 

 \begin{equation}
 \label{eq:mode_ampl_lin}
     |a|^2 = \frac{\gamma_r|f|^2}{\gamma^2 + \Delta\omega ^2},
 \end{equation}
where $\Delta \omega=\omega_0-\omega$ is the detuning between the pumping field and the mode frequency. The mode intensity equals the total energy stored inside the resonator $W=|a|^2$. Thus, the absorbed power  $P_{\text{abs}}=2\gamma_{\text{nr}}|a|^2$  also has resonant Lorentz spectral profile and at the resonance reaches its maximal value

$$
P_{\text{abs}}|_{\omega=\omega_0}=\frac{2\gnr \gr |f|^2}{(\gr+\gnr)^2}.
$$
In these terms, the critical coupling condition manifests itself in equal radiative and non-radiative losses $\gr=\gnr$. Indeed, once this condition is fulfilled,  the absorbed power tends to its maximal value of $P_{\text{abs}}\to |f|^2/2$.      


 \subsection{Bistability  of nonlinear resonator}
   \label{sec:Bistab}
Due to the thermorefractive effect, both real and imaginary parts of the refractive index of the material may start to depend on temperature resulting in correspondent dispersive and absorptive Kerr-type nonlinearities. One of the main consequences is the appearance of a bistability regime~\cite{Lugiato1983} when the system may have several stationary states. In this section, we will discuss this behavior in more detail as it is directly related to nonlinear critical coupling conditions and optimized optical heating. {It is }{worth mentioning, that we neglect the thermal expansion effect as it gives a much smaller contribution to the resonance shift in comparison with the thermooptical effect (see Table 2~\cite{zograf2021all})}.    

Within a single-mode approximation, the thermooptical Kerr nonlinearity can be accounted for in the first order of perturbation theory through the shift of frequency spectral position $\omega_0\to\omega_0 -\alpha |a|^2$ and varied non-radiative losses $\gnr\to\gnr +\beta | a|^2$. Here, we assume that both nonlinear coefficients are positive $\alpha,\beta>0$ which results in the redshift of the resonance and increase of losses with temperature increase. The exact expressions are presented in the following section~\ref{sec:alpha_beta}. With an account for this the equation for the stationary mode amplitude Eq.~\eqref{eq:mode_ampl_lin} will be modified:

\begin{align}
 \label{eq:mode_ampl_nonlin}
     |a|^2 = \frac{\gamma_r|f|^2}{(\gamma+\beta|a|^2)^2 + \left(\Delta\omega-\alpha |a|^2\right)^2}\to \nonumber \\
          y = \frac{y_0}{(1+b y)^2 + \left(\Delta\tilde\omega-y\right)^2}.
\end{align}

Here, we { introduce} dimensionless parameters {} for the mode intensity $y=\alpha |a|^2/\gamma$, the external pump intensity $y_0=\alpha \gr|f|^2/\gamma^3$, the relative nonlinear coefficient $b=\beta/\alpha$, and normalized frequency detuning $\Delta \tilde \omega= \Delta\omega/\gamma = (\omega_0 - \omega)/\gamma$. One can notice that in the system there are only three independent parameters: the relative nonlinear coefficient $b$, normalized detuning frequency $\Delta \tilde \omega$, and normalized pump intensity $y_0$. Interestingly, to analyze the pump intensity $y_0=\Delta\omega_{\text{eff}}/\gamma$ which has the form of the ratio of the effective spectral shift of the resonance under the external pump $\Delta\omega_{\text{eff}}=\alpha |f|^2/\gamma^2$ over the spectral width of the resonance $\gamma$.  Thus, to increase $y_0$ one can either increase the pump field intensity, decrease the total losses in the system, or increase the nonlinear coefficient making the system more sensitive to heating.  Alternatively, the equation can be rewritten in a more compact form

\begin{align}
\label{eq:mode_ampl_nonlin2}
        y = \frac{y_{\text{max}}}{1+\left(y-y_s\right)^2/\Gamma^2},  &\quad \text{\ where}\\
 y_{\text{max}}=\dfrac{y_0(b^2+1)}{(b\dwt+1)^2}, \quad  y_s=\dfrac{\Delta \tilde{\omega}-b}{1+b^2}, & \quad \Gamma=\dfrac{1+b\Delta\tilde{\omega}}{1+b^2}.\nonumber
\end{align} 

Eq.~\eqref{eq:mode_ampl_nonlin}-\eqref{eq:mode_ampl_nonlin2} is the central equation describing the state of the nonlinear resonator and its solution $y$ describes  the  intensity of the field inside the resonator. However, this is a third-order algebraic equation and it may have not a unique solution but also three solutions that correspond to the bistability regime~\cite{Lugiato1983}. Indeed, the dependence of the mode intensity $y$ on the pump intensity $y_0$ is shown in Fig.~\ref{fig:Model_Bistab} (a) for different values of the detuning parameter demonstrating highly nonlinear dependence. Moreover, for the detuning, $\Delta \tilde \omega=3$ one can observe the hysteresis behavior. The loop region corresponds to three different solutions and only two of them are stable at the upper and lower parts of the s-type curve~\cite{Lugiato1983} (see Fig.~\ref{fig:Model_Bistab} (a)). For the larger values of detuning $\dwt=5$ the hysteresis will be observed at larger pump intensities $y_0$ not shown in the plot.

Careful analysis of Eq.~\ref{eq:mode_ampl_nonlin} shows that there exists a set of parameters that provide the bistability condition. To illustrate that, we plot a map  showing  the mode intensity $y$ dependence on pump intensity $y_0$ and detuning $\dwt$ (see Fig.~\ref{fig:Model_Bistab} (b)) for fixed value of the relative  nonlinear  coefficient $b=0.05$. The color shows the mode intensity normalized over the maximal value $y_{\text{max}}$ in Eq.~\eqref{eq:mode_ampl_nonlin2} and discussed below. The hatched area denotes the region of bistability and the upper branch of the solution is depicted on the map. One can see the abrupt drop of the intensity at the right edge of the bistability region corresponding to the drop from the upper to lower branch of the solution (point $B$ in the figure). The points $A$ and $B$ show the { intensities corresponding to the} switching between the upper and lower branches of the solution (see also Fig.~\ref{fig:Model_Bistab} (a)).     
  
Further analysis shows that the bistability region strongly depends on $b$, which is the ratio of the imaginary and real parts of thermorefractive constant. Critical point $O$ which denotes the appearance of the bistability region moves upwards on the $(\dwt, y_0)$ map with the increase of $b$ as illustrated in Fig.~\ref{fig:Model_Bistab} (c) where the bistability regions are shown for varied parameters $b$. It turns out that for $b=0$, which corresponds to an absence of thermal dependence of ohmic losses, the bistability can appear at the smallest value of the intensity $y_0^*=8\sqrt{3}/9$ and detuning $\dwt^*=\sqrt{3}$.     With the increase of  $b$ the bistability region moves upwards and shrinks in the spectral width.  

Such behavior has a clear physical explanation. Large $b$ means that the losses in the system rapidly increase with the amplitude of mode. To maintain bistability one needs to decrease the losses in the 'cold' system $\gamma$, which means to increase the pump intensity $y_0$. However, what is less obvious, that the bistability exists only for  values of $b$ smaller the $b<b_{\text{cr}}=1/\sqrt{3}$. Indeed, once $b\to b_{\text{cr}}$ the point $O$ goes to infinity along the black dashed line in Fig.~\ref{fig:Model_Bistab}(c). Thus, for a large enough absorptive nonlinear coefficient $\beta>\alpha/\sqrt{3}$ the bistability can not be observed. {
One can interpret the existence of {the} critical value $b_{\text{cr}}$ in the following manner: at large $\beta/\alpha$, the spectral width of the resonance increases with temperature much faster than the resonant spectral position shifts, and the bistability condition {(}i.e. the shift of the peak should be larger than its width{)} simply can not be achieved. This does not allow to achieve a bistability regime in a single-mode nonlinear resonator (for more details see Supplementary Information S1).

Finally, { it is important to discuss} the hysteresis loop parameters such as hysteresis width $W_{\text{h}}$ and height $H_{\text{h}}$, which is the $x$-axis and $y$-axis distance between the turning points ($A$ and $B$) in Fig.~\ref{fig:Model_Bistab}(a),{ respectively}. In the case of purely dispersive nonlinearity $b=0$, the parameters can be immediately derived in a simple form (see details in Supplementary Information{, Section} S1):
\begin{equation}
\begin{aligned}
    H_{\text{h}} &= y(A)-y(B)=\frac{2}{3}\sqrt{\dwt^2 - 3};\\
    W_{\text{h}} &= y_0(B)-y_0(A)=\frac{ H_{\text{h}}^3}{2}.
\end{aligned}
\end{equation} 

One can see that the height and width of the hysteresis loop appear to be dependent on the detuning frequency and are mutually dependent. For large frequency detuning values $\dwt \gg 1$, these quantities {} behave as $H_{\text{h}}\sim\dwt$ and $W_{\text{h}} \sim \dwt^3$, whereas for close to the critical point values $\dwt = \dwt^* + \delta$ { they have the following dependencies on the detuning:} $H_{\text{h}}\sim\delta^{1/2}$ and $W_{\text{h}}\sim\delta^{3/2}$. Consequently, for small values of detuning near the critical point hysteresis height increases more rapidly than its width. Thus, for potential {optical} switching { applications}, it is more prospective to work closer to {the} critical frequency $\dwt^*$, where the height of the hysteresis loop is high, while the width is small providing a stronger amplitude difference between the stationary states at lower switching intensities. The graphical plots of these parameters along with additional details on $b > 0$ case are provided in Supplementary Information materials, Section{, Section} S1.}


\subsection{Nonlinear critical coupling}

Till now, we have discussed the possible states of the resonator with thermally induced Kerr nonlinearity.  The maximal possible mode intensity is provided by the nonlinear critical coupling condition based on the straightforward analysis of Eq.~\eqref{eq:mode_ampl_nonlin}. The Lorentz-type of right-hand side of Eq.~\ref{eq:mode_ampl_nonlin2} ensures that the  value $y$ can not be greater than $y\leq y_{\text{max}}$. Moreover, it turns out that there are a certain set of parameters $(y_0,\dwt,b)$ for which the maximal value $y_{\text{max}}$  is reached and they are given by the equation
\begin{align}
\label{eq:nonl_critical}
(\dwt - b)(b\dwt +1)^2=y_0(b^2+1)^2.
\end{align}

This equation defines the {\it nonlinear critical coupling} conditions maximizing the mode intensity $y$. In the first-order perturbation, one can obtain the proper detuning frequency $\dwt\approx y_0+b$ provides the critical coupling. This condition means that to reach the maximal mode intensity one should have particular detuning between the pump frequency and the "cold" resonant frequency. The stronger the pump intensity the larger should be the detuning. Now, once the $b$-coefficient becomes stronger the higher-order dependence appears providing that $y_0\sim\dwt^3$. 
 
The critical coupling regime is seen in Fig.~\ref{fig:Model_Bistab} (b) where the normalized mode intensity is plotted as the function of $\dwt$ and $y_0$. One can see that the maximal value is reached along the black dashed line, which is obtained as the solution of the nonlinear critical coupling equation Eq.~\eqref{eq:nonl_critical}.  One can see that the critical coupling is reached close to the edge of the bistability region at the upper branch of the solution. So, to achieve critical coupling one needs to drive the system in the bistable state at the upper branch.     

{The nonlinear critical coupling condition formulated in Eq.~\eqref{eq:nonl_critical} in dimensionless units can be reformulated in terms of radiative and non-radiative losses, which is convenient for the designing the optical resonators}. The total absorption in the regime of maximal heating when $y=y_{\text{max}}$ will have a for similar to the classical one with slight modification

\begin{align}
P_{\text{abs}}=\frac{2(\gnr+b\dw) \gr |f|^2(1+b^2)}{(\gr+\gnr+b\dw)^2}.
\end{align}
 Now for any fixed value of $\gr$ the maximal absorption will be observed at 
\begin{align} 
\gr=\gnr+b\dw,
\label{eq:crit_coupl2}
\end{align}
{which is an analog of the classical critical coupling condition and has a clear physical meaning, at least in the first order perturbation with respect to $b$}: i) the system in its "cold" state should be out of the {\it linear} critical coupling condition; ii) once the resonator is pumped with the particular  intensity and at spectral detuning in accordance to  Eq.~\eqref{eq:crit_coupl2}, in the final "hot" stationary state the radiative and non-radiative losses should be balanced. { Indeed, expanding Eq.~\eqref{eq:nonl_critical} in the series with respect to $b\ll1$, one obtains that $\dwt=y_{0}+O(b)=y_{\text{max}}+O(b)$ and $b\dw\approx\beta |a|^2_{\text{max}}$. Thus, the second term in Eq.~\eqref{eq:crit_coupl2} simply corresponds to the added losses due to the heating of the resonator.}

 \subsection{Nonlinear coefficients $\alpha$ and $\beta$}
 \label{sec:alpha_beta}
The behavior of the nonlinear resonator strongly depends on the values of dispersive and absorptive nonlinear coefficients $\alpha$ and $\beta$  which provide the nonlinear coupling.   Their amplitude is fully defined by the resonant mode properties and thermorefractive coefficients expressing the thermal origin of nonlinear coupling. One can derive it starting from very general considerations of the absorbed power:
 
 \begin{equation}
    P_{\text{abs}} = \frac{\omega}{2}\varepsilon_0\varepsilon''\int_{V_{NP}}|\mathbf{E(r)}|^2dV, \label{eq: p_abs}
\end{equation}
where $\varepsilon_0$ is the permittivity of vacuum, $\varepsilon''$ is the imaginary part of the complex dielectric permittivity $\varepsilon = \varepsilon' + i\varepsilon''$, $\omega$ is the excitation frequency and $\mathbf{E(r)}$ is the electric field amplitude inside of the nanoparticle, and the integration is taken over the volume of the particle. We can introduce electric field through the mode amplitude as $\mathbf{E(r)} = a\mathbf{M(r)}/\sqrt{\varepsilon_0}$, where $\mathbf{M(r)}$ is the normalized eigenmode field distribution~\cite{doost2014resonant}. For high-Q modes this normalization can be approximated by the expression  $\int_{V_{NP}}\varepsilon'|\mathbf{M(r)}|^2dV = 1$ . The thermorefractive effect in the linear  approximation can be introduced as follows: 
\begin{equation}
    \begin{aligned}
        &n = n_0 + n_1\Delta T,\ \  k = k_0 + k_1\Delta T,
    \end{aligned}\label{eq: refr_ind}
\end{equation}
where $n_0$ and $k_0$ are the initial values of complex refractive index at room temperature $T = 298$K, $n_1 = dn/dT|_{T = 298\mathrm{K}}$ and $k_1 = dk/dT|_{T = 298\mathrm{K}}$ are the linear thermorefracitve coefficients. The imaginary part of the dielectric permittivity $\varepsilon_0''$  also increases linearly with temperature and in the first order approximation  we have then $\varepsilon'' = 2nk \approx2n_0k_0+ 2(n_0k_1 + k_0n_1)\Delta T = \varepsilon''_0 + \varepsilon_1''\Delta T$ and expression \eqref{eq: p_abs} could be rewritten as:
\begin{equation}
     P_{\text{abs}} = \frac{\omega}{2}\frac{\varepsilon''_0}{\varepsilon'}|a|^2 + \frac{\omega}{2}\frac{\varepsilon''_1}{\varepsilon'}\Delta T|a|^2. \label{eq: p_abs_new}
\end{equation}
Finally, we relate the temperature with the    absorbed electromagnetic power $\Delta T = \kappa P_{\text{abs}}$ through the linear coefficient $\kappa$ similarly to Eq.~\eqref{eq:temp}. This coefficient depends on the particular geometry of the system and can be found with help of exact numerical simulation, for instance. With this, the temperature can be expressed from Eq.~\eqref{eq: p_abs_new} 
\begin{equation}
    \Delta T = \frac{\Delta\tilde{ T}}{1-\Delta\tilde{T} /\Delta T_{\mathrm{c}}} \approx \Delta\tilde{T},
    \label{eq: temp_increase}
\end{equation}
where $\Delta\tilde{T} = \kappa\omega\varepsilon_0''|a|^2/2\varepsilon' = \kappa\omega n_0k_0|a|^2/\varepsilon'$ is proportional to the mode intensity, and $\Delta T_{\mathrm{c}} = \varepsilon_0''/\varepsilon_1'' = n_0k_0/(n_0k_1 + k_0n_1)$ is  critical temperature increase. One can see that the temperature increase by itself has a nonlinear relation with the mode amplitude. However, as far as the temperature increase is smaller than the critical one $\Delta\tilde{T}/\Delta T_{\mathrm{c}}\ll 1$ the connection can be linearized. 

For a dielectric resonator the  frequency shift is governed by the change of the refractive index, thus providing  
$\Delta\omega_0/\omega_0 = -\Delta n/n$. Recalling introduced mode spectral shift from Sec.~\ref{sec:Bistab} and temperature dependence of the complex refractive index \eqref{eq: refr_ind}, we obtain
\begin{equation}
    \Delta T = \frac{n_0\alpha}{n_1\omega_0}|a|^2, \label{eq: temp_increase2}
\end{equation}
where $\omega_0$ is unperturbed resonator eigenfrequency. Combination of expressions \ref{eq: temp_increase} and \ref{eq: temp_increase2} gives us the value of thermooptical coefficient $\alpha$:
\begin{equation}
    \alpha \approx \frac{n_1\omega_0}{n_0}\cdot\Delta \tilde{T}/{|a|^2} \approx \frac{\kappa\omega_0^2k_0n_1}{\varepsilon'} \label{eq: alpha},
\end{equation}
where the last approximation is made for small values of frequency detuning $\Delta\omega/\omega_0 \ll 1$.

Similarly, we can derive the absorptive nonlinear coefficient $\beta$.  From the amplitude non-radiative losses rate definition:
\begin{equation}
    \gamma_{\mathrm{nr}} = \frac{P_{\text{abs}}}{2W} = \frac{\omega}{4}\frac{\varepsilon''(T)}{\varepsilon'} = \frac{\omega}{4\varepsilon'}\left(\varepsilon_0'' + \varepsilon_1''\Delta T\right). 
\end{equation}
Using expression~\eqref{eq: temp_increase2} for the relation between temperature increase inside of the nanoparticle and mode amplitude, we obtain the nonlinear coefficient $\beta$:
\begin{equation}
    \beta = \frac{\omega}{4}\frac{\varepsilon_1''}{\varepsilon'}\frac{n_0\alpha}{n_1\omega_0}\approx
    \frac{\kappa\omega_0^2n_0k_0\left(n_0k_1+n_1k_0\right)}{2\varepsilon'^2} \label{eq: beta},
\end{equation}
which is true if we work in the vicinity of an unperturbed eigenfrequency position. Surprisingly, the relative  thermooptical parameter $b$, which governs the general behavior of the nonlinear resonator, has the following form:
\begin{equation}
    b = \frac{\beta}{\alpha} = \frac{n_0\left(n_0k_1+n_1k_0\right)}{2n_1\varepsilon'}=\dfrac{n_0}{2n_1}\dfrac{\varepsilon_1''}{2\varepsilon_0'}, \label{eq: thermoopt_par}
\end{equation}
which does not depend on the resonant characteristics of the nanoparticles but only on the material properties and thermorefractive characteristics.
\section{Quasi-BIC nanoresonator heating}

In this section, we implement the developed approach for optimizing the optical heating of a single nanoresonator. In strong contrast to nanoplasmonic designs, we aim at semiconductor materials, which demonstrate strong thermooptical nonlinear coefficient~\cite{zograf2021all} and also can demonstrate high-Q resonance which provides strong thermooptical reconfiguration under moderate laser intensities. Moreover, the semiconductor structures allow for precise tuning of losses in a wide spectral range by doping the material with free carriers. Together with the fine engineering of radiative losses with the concept of quasi-BIC modes~\cite{koshelev2020subwavelength, bogdanov2019bound} one can achieve high efficiency of optical heating of a single wavelength scale resonator.

\subsection{Radiative and non-radiative losses optimization }

\begin{figure}[h!]
\includegraphics[width=\textwidth]{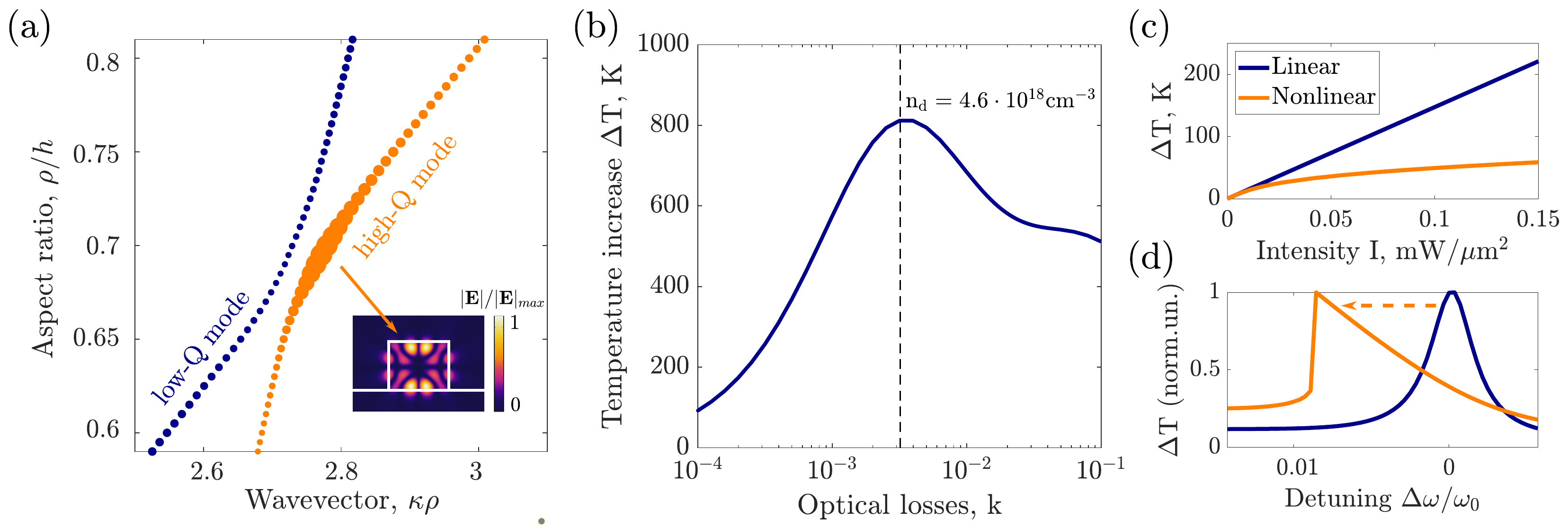}
\caption{a) The cylinder eigenmodes positions depending on the dimensionless wavevector $\kappa\rho = 2\pi\rho/\lambda$, where $\lambda$ -- mode free-space wavelength, $\rho$ -- cylinder radius, and on the dimensionless size parameter $\rho/h$, where $h$ is the cylinder height. The size of the dots is proportional to the eigenmode quality factor. The inset shows a normalized electric field for quasi-BIC mode with quality factor Q = 615. b) Simulated temperature increase in a cylinder with quasi-BIC mode as the function of the imaginary part of refractive index k responsible for optical losses of the system. Dashed line shows the optimal value of the losses which corresponds to the concentration of dopants $n_d = 4.6\cdot 10^{18} \, \mathrm{cm}^{-3}$. c) Comparison of linear (without thermorefraction) and nonlinear (with thermorefraction) regimes of heating the nanoparticle for zero detuning $\delta\omega = 0$. d) Spectral shift of heating characteristic from linear (orange) to nonlinear (blue) one for thermorefractive system. The incident intensity is I = 0.275 $\mathrm{mW/\mu m^2}$ \label{fig:liner_heating}}
\end{figure}

Quasi-BIC states are characterized by the destructive interference of radiation in the far-field zone and, hence, efficient localization of electromagnetic energy in the resonator volume~\cite{bogdanov2019bound}. The high-Q states in nanocylinders can be excited with polarized vector beam~\cite{koshelev2020subwavelength} enhancing light-matter interaction. Full-wave numerical simulations allowed us to locate the spectral position of high-Q modes silicon nanocylinders on quartz substrates (see Fig.~\ref{fig:liner_heating} (a) inset).  The modes of cylindrical resonators are classified by their azimuthal number $m$~\cite{Gladyshev2020}, which indicates the symmetry of the electromagnetic field $E, H \sim e^{im\varphi}$, where $\varphi$ is the azimuthal angle to cylinder axes. We investigate the azimuthally symmetrical ($m = 0$) eigenmodes matching the symmetry of the incident azimuthal vector beam (see Fig.~\ref{fig:general}). The map showing the eigenmodes spectral position as a function of the aspect ratio of the cylinder is shown in  Fig.~\ref{fig:liner_heating} (a).  One can see the formation of a high-Q state due to coupling between two modes in the vicinity of the anticrossing region. Fixing then the incident wavelength $\lambda = 1400$ nm we obtain geometrical parameters for the declared quasi-BIC state which are radius $\rho$~=~1238~nm ad height $h$~=~885~nm. 


Next, we have optimized the non-radiative losses to get close to the critical coupling condition. The doping of semiconductors provides an additional degree of freedom for precise non-radiative losses control. We choose silicon for the resonator material since it possesses zero optical losses in the near-IR region~\cite{green2008self} which gives an opportunity to finely tune the losses in the wide range via modulation of dopant concentration. 

For chosen geometry of the resonator, we perform rigorous optimization for non-radiative losses by simulating in numerical commercial software COMSOL Multiphysics temperature increase inside of the nanoparticle excited by the azimuthal vector beam~\cite{novotny2012principles} Fig.\ref{fig:liner_heating} (b). Analysis of the temperature as a function of the imaginary part of the refractive index $k$ responsible for the non-radiative losses rate in the system allows identifying the critical coupling value of optical losses $k = 0.0031$. The  estimated  concentration of dopants providing this condition according to the Drude model is expressed as:
\begin{equation}
\Delta\mathrm{Im}\varepsilon(n_d)=\frac{{\omega_{p}}^2(n_d)\tau_{\mathrm{e}}}{\omega(1+\omega^2\tau_{\mathrm{e}}^2)} = 2nk,
\end{equation}
where $\omega$ is the frequency of the incident radiation; $\tau_{\mathrm{e}}$ = 1 fs is the electron momenta relaxation time; plasmonic frequency $\omega_p(n_d) = (n_de^2/\varepsilon_0m_{\mathrm{eff}})^{1/2}$; $m_{\mathrm{eff}}=0.18m_{\mathrm{e}}$ is the effective mass of electrons in the conduction band of c-Si~\cite{sokolowski2000generation}; $\varepsilon_0$ is the permittivity of vacuum, and $e$ is the elementary charge. Assuming that the real part of the refractive index ${n}$ does not change significantly with the free carrier concentration increase~\cite{makarov2015tuning}, we immediately obtain donors concentration $n_d = 4.6\cdot 10^{18} \, \mathrm{cm}^{-3}$ corresponding to the optimal value of $k=0.0031$ Extracted total loss rate for the resonator with optimized parameters is then
$\gamma = \gr + \gnr = 2.4\cdot 10^{12}$~1/s.

\subsection{Numerical modeling of optical heating}

We have performed the simulations on the optical heating of the designed nanoresonators demonstrating a huge temperature increase $\Delta T = 815$K for relatively low incident power flux $I = 0.55 \mathrm{mW/\mu m^2}$ in the linear regime when the thermooptical effects are omitted. However, once the nonlinearity of the system is taken into account,   heating is significantly suppressed under the resonant excitation (compare linear and nonlinear regimes in  Fig.~\ref{fig:liner_heating} (c)). Such a noticeable deviation from the linear trend is associated with the eigenfrequency spectral shift shown in \ref{fig:liner_heating} (d). Thus, for maximizing heating efficiency in a nonlinear thermorefractive system one needs to obtain the optimal condition concerning both excitation frequency and incident intensity by the basic theory described in Sec.\ref{sec:the_model}.

In our design of the nanoresonator tuned for IR-region, the relative absorptive nonlinear coefficient $b$ appears to be negligibly low $b \approx 0$. Indeed, in the near-IR region crystalline silicon has zero intrinsic optical losses and therefore non-radiative losses in the system are only defined by the concentration of dopants. Since typical values of donors ionization energy are less than thermal energy for the room temperature $E\sim kT$, we suppose them to be fully ionized at $T = 298$K~\cite{kohn1955theory}. Consequently, temperature increase does not influence optical losses and imaginary part of the thermorefractive coefficicent can be set to zero $k_1 = dk/dT|_{T = 298\mathrm{K}} = 0$. At the same time, the real part of the thermooptical coefficient    at the wavelength $\lambda = 1400$ nm equals to  $n_1 = 2\cdot 10^{-4}$ 1/K~\cite{frey2006temperature} which from Eq.~\eqref{eq: alpha}-\eqref{eq: beta} results into $\alpha = 3.25\cdot10^{28}\mathrm{J^{-1}s^{-1}}$ and $\beta = 1.44\cdot10^{25}\mathrm{J^{-1}s^{-1}}$ nonlinear coefficicents. 

\begin{figure*}[h!]
\includegraphics[width=1\textwidth]{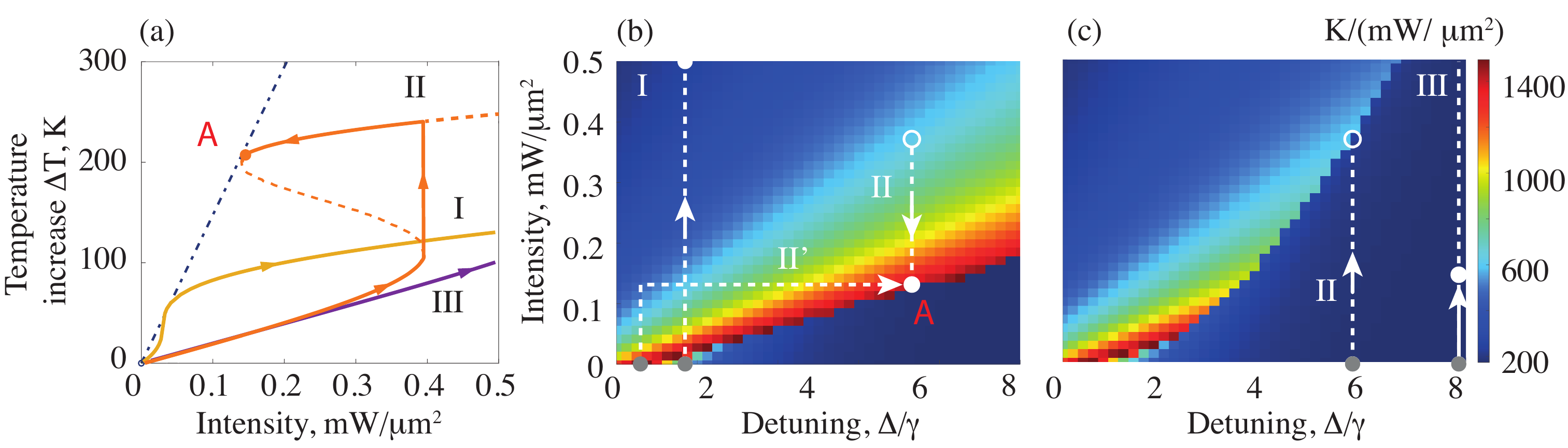}
\caption{(a) The dependence of the temperature on the pump intensity for different detuning parameters $\dw/\gamma=1.5$ (I),$\dw/\gamma=6$ (II), $\dw/\gamma=8$ (III).   (b,c) Heating efficiency maps $ \Delta T/I_0$ of the doped silicon ($n_d = 4.6\cdot 10^{18} \, \mathrm{cm}^{-3}$) cylinder with radius $\rho$ = 1238 nm and height $h$ = 885 nm, as the function of pump intensity $I_0$ and detuning factor $\Delta\tilde{\omega} = \Delta\omega/\gamma$ for (b) upper and (c) lower  branches of the solution.   The difference in the maps denotes hysteresis region.}
\label{fig:Numerical_map}
\end{figure*}

With the given material parameters, we perform full-wave simulations coupled with the heat transfer module in commercial software COMSOL Multiphysics.   The thermooptical coupling provides the nonlinear response of the simulated system which may initiate computational difficulties once the bistability regime is reached.  In this case, the final state of the iterative numerical solution depends on the initial solutions guess whether the solution is located on the upper, lower, or unstable part of the s-curve shown in Fig.~\ref{fig:Numerical_map} (a). For that, we were carefully choosing the initial guess and used the obtained solutions as the initial guess for the next set of parameters. The dependence of the heating efficiency denoting the temperature increase per unit incident power $\Delta T/I_0$ is shown in Fig.~\ref{fig:Numerical_map} (b) and (c) in full analog with the plots shown in Fig.~\ref{fig:Model_Bistab} (b) for a toy model.  Figure~\ref{fig:Numerical_map} (b) and (c) shows the heating efficiency at the upper and lower branch of the solution correspondingly. One can see that the maximal efficiency of heating is indeed reached at the upper branch by the basic theory discussed in Sec.~\ref{sec:Bistab}. {To reach this solution in the simulations, one can simply start with the high intensity of the laser pump and high temperature initial guess and then gradually decrease the pump power. However, this is not very physical from the experimental point of view, when normally the excitation of the systems starts from the cold regime. In this view, we {show} several  ``trajectories''  in the parameters space (see Fig.~\ref{fig:Model_Bistab}), which { are} tested in numerical simulations. The trajectory I stands for the increase of the incident power at the frequency detuning where there is no bistability region. For the trajectory II the detuning frequency is larger and with the increase of the intensity one starts at the lower branch of the solution (see Fig.\ref{fig:Model_Bistab}) and then by a gradual increase of the pump intensity reaches the upper branch, after that with the decreasing of the temperature one can reach the maximal heating efficiency point $A$. However, such a route leads to necessary  overheating of the structure: reaching the optimal heating point one first need to jump on the upper steady-state branch at higher temperature and then get to the critical point A (see see Fig.\ref{fig:Model_Bistab} (b) and trajectory II there). Thus, alternatively, one can choose the trajectory II' where one first increases the pump at small detuning and then gradually increases the detuning at constant power reaching again point $A$. Finally, the shown trajectory III does not reach the supper branch and the heating efficiency is kept at a very low level. }


It is also worth noting that at the point of maximal heating efficiency in the nonlinear regime, its value becomes equal to the heating efficiency in the linear regime (see the linear dash-dotted line in Fig.~\ref{fig:Numerical_map} (a)), which also matches the basic model described in Sec.~\ref{sec:the_model}. However, that occurs only in the case of small dissipative nonlinearity $b\approx 0$.

\section{Discussion and Conclusion}

The main idea of this paper is to reveal the key aspects of optical heating of a semiconductor resonator: i)  once the resonator is heated in the CW regime, the efficiency of heating immediately drops due to thermooptical reconfiguration of the resonator; ii) one can reach high efficiency only the heating is accompanied with proper spectral detuning of the pumping laser from the 'cold' resonance of the mode; iii) The maximal heating efficiency is reached in the bistability regime which appears under certain conditions on the pumping intensity, thermooptical coefficients, and of the resonant mode.

However, despite the discussed design being aimed at the near-IR range, where the intrinsic losses of silicon are vanishing, one can tune the proposed design closer to the visible spectra and balance the radiative losses with the intrinsic losses of silicon, which occur at around $\lambda$ = 865 nm. Here, the imaginary part of thermorefractive coefficient is non-zero $k_1= 8.5\cdot 10^{-6}$~1/K~\cite{jellison1994optical}, which provides the relative nonlinear coefficient $b=0.015$.  

It is also worth mentioning, that the suggested effect of nonlinear critical coupling can be observed not only in the CW regime, but also under the pulsed excitation once the pulse duration is long enough so that the equilibrium temperature is achieved, i.e. for nanosecond laser pulses. In that case, the quasi-CW regime can be considered and chirped laser pulses could be used to achieve the efficient heating of nanoresonators. Alternatively, the heating effects under the short pico- and femtosecond pulses excitation requires more  complex models based on the analysis of  nonequilibrium dynamics of carriers~\cite{zograf2021all}. 
 
Finally, we would like to provide a brief comparison of the proposed nanophotonic design in terms of heating efficiency to existing analogs of nanoscale optical heaters. Indeed, Fig.~\ref{fig:Numerical_map} depicts the map of heating efficiencies as a function of pump intensity and the detuning factor. The maximum values of heating efficiencies reached 1400~K/(mW/$\mu$m$^2$), which is the highest value of heating efficiencies for the structures with direct thermal contact with substrates. For a single c-Si nanodisk laser heating on a substrate at magnetic dipole and quadrupole modes the efficiency reaches 150~K/(mW/$\mu$m$^2$)~\cite{zograf2018local}, and for a c-Si sphere is about 300~K/(mW/$\mu$m$^2$)~\cite{zograf2017resonant} with considerably less thermal contact with the substrate than the disk has. The latter results were obtained experimentally, however, the best theoretical values reach 750~K/(mW/$\mu$m$^2$) for complex semiconductor structures with quasi-infinite c-Si nanorod covered with a-Si film~\cite{Danesi2018} and one of the most promising nanostructures for light-to-heat conversion is golden doughnut {supporting efficiency values} up to 230~K/(mW/$\mu$m$^2$) in aqueous media. {Nevertheless}, it remains questionable of fabrication, experimental feasibility, and temperature detection possibility~\cite{gonzalez2022gold}. {On the other hand, thermal nonlinearity {driven} by pulsed laser heating is also a rather promising approach{, but} the experimentally demonstrated efficiency for c-Si nanocubes reached only 30~K/(mW/$\mu$m$^2$)~\cite{duh2020giant} mostly due to oil immersion, which drastically increases the thermal conductivity of the nanoresonator's surrounding medium.}

In conclusion, we have developed a new simple theoretical approach to optimization of the resonator optical heating in the nonlinear regime. The proposed design based on the super-cavity mode in doped silicon cylindrical particles has allowed for efficient light-to-heat conversion when the initial spectrum of incident laser is detuned from the initial spectral position of the resonance. Moreover, we have revealed a bistability regime in the optical heating at an intensity around 1~mW/$\mu$m$^2$. 
Our results are also helpful for resolving the thermal challenges for all-dielectric resonator-based photonic devices~\cite{padmaraju2014resolving,wang2020athermal}, Raman microlasers~\cite{agarwal2019nanocavity,zograf2020stimulated}, and nanoscale photo-thermal chemistry and sensing~\cite{caldarola2015non, regmi2016all, milichko2018metal}.
As an outlook, we believe that the developed bistability approach is quite universal and can be further applied not only for various types of nonlinearities based on Kerr effect~\cite{shcherbakov2015ultrafast, grinblat2020efficient}, electron-hole plasma generation~\cite{sokolowski2000generation,makarov2015tuning, sinev2021observation} and excitonic effects~\cite{gibbs1979optical, gibbs1982room,masharin2022polaron}.

\begin{acknowledgement}

\end{acknowledgement}

\begin{funding}
This research was supported by Priority 2030 Federal Academic Leadership Program and by the Ministry of Science and Higher Education of the Russian Federation (Project 075-15-2021-589).
\end{funding}

\bibliography{bibliography2.bib}
\end{document}